\documentclass{acm_proc_article-sp}
\include{epsf}

\newcommand{\commentout}[1]{}

\newenvironment{ctable*}
{\begin{table*}[htpb]\begin{center}}{\end{center}\end{table*}}

\newtheorem{theorem}{Theorem}[section]
\newtheorem{example}{Example}[section]
\newtheorem{proposition}[theorem]{Proposition}
\newtheorem{corollary}[theorem]{Corollary}
\newtheorem{lemma}[theorem]{Lemma} 

\newtheorem{definitionx}[theorem]{Definition}

\newenvironment{definition}{\begin{definitionx}\rm}{\end{definitionx}}

\newcommand{\Increase}{\mbox{{\it Increase}}}
\newcommand{\bcast}{\mbox{{\it bcast}}}
\newcommand{\send}{\mbox{{\it send}}}
\newcommand{\recv}{\mbox{{\it recv}}}
\newcommand{\dir}{\mbox{{\it dir}}}
\newcommand{\join}{\mbox{{\it join}}}
\newcommand{\leave}{\mbox{{\it leave}}}
\newcommand{\angleChange}{\mbox{{\it aChange}}}
\newcommand{\cov}{\mbox{{\it cover}}}
\newcommand{\gap}{\mbox{{\it gap}}}
\newcommand{\cone}{\mbox{{\it cone}}}
\newcommand{\cir}{\mbox{{\it circ}}}
\newcommand{\rad}{\mbox{{\it rad}}}
\newcommand{\inter}{\cap}

\begin{document}


\newcommand{\remove}[1]{}
\newcommand{\addspace}{}

\title{Analysis of a Cone-Based Distributed Topology Control Algorithm for Wireless
Multi-hop Networks}

\author{
\setlength\tabcolsep{3\tabcolsep}
\begin{tabular}{cc}
Li  Li & Joseph Y. Halpern   \\
{\affaddr Department of Computer Science} & {\affaddr Department of Computer Science} \\
{\affaddr Cornell University}             & {\affaddr Cornell University} \\
{\tt lili@cs.cornell.edu}            & {\tt halpern@cs.cornell.edu}  \\
\end{tabular}
\\
\\
\setlength\tabcolsep{3\tabcolsep}
\begin{tabular}{ccc}
 Paramvir Bahl       &  Yi-Min Wang &  Roger Wattenhofer  \\
{\affaddr Microsoft Research}    & {\affaddr Microsoft Research} & {\affaddr Microsoft Research} \\
{\tt bahl@microsoft.com} & {\tt ymwang@microsoft.com} & {\tt rogerwa@microsoft.com} 
\end{tabular}
}

\maketitle


\begin{abstract}

The topology of a wireless multi-hop network can be controlled by
varying the transmission power at each node.
In this paper, we give a detailed analysis of a 
cone-based distributed topology control algorithm.
This algorithm, introduced in \cite{ourInfocom01}, does not assume that
nodes have GPS information available; rather it depends only on 
directional information.
Roughly speaking, the basic idea of the algorithm is that a node $u$
transmits with the minimum power $p_{u,\alpha}$
required to ensure that in every 
cone of degree $\alpha$ around $u$,
there is some node that $u$ can reach with power $p_{u,\alpha}$.  
We show that taking $\alpha = 5\pi/6$ is a necessary and sufficient
condition to guarantee that network connectivity is preserved.  
More precisely, 
if there is a path from 
$s$ to $t$ when every node communicates at
maximum power
then, if $\alpha \le 5\pi/6$,
there is still a path in the smallest symmetric graph $G_\alpha$
containing all edges $(u,v)$ such  that $u$ can communicate with $v$
using power $p_{u,\alpha}$.  On the other hand,
if $\alpha > 5\pi/6$,  connectivity is not necessarily preserved.
We also propose a set of optimizations that further reduce power
consumption and prove that they retain network connectivity.
Dynamic reconfiguration in the presence of failures and mobility is also
discussed.
Simulation results are presented to demonstrate the effectiveness
of the algorithm and the optimizations.
\end{abstract}


\section{Introduction}
\label{sec-intro}
Multi-hop wireless networks, such as radio networks \cite{kahn77},
ad-hoc networks \cite{Perkbook01} and sensor 
networks \cite{Clare99,Pottie00}, 
are networks where communication between
two nodes may go through multiple consecutive wireless links. 
Unlike wired networks, which typically have a fixed 
network topology
(except in case of failures), each node in a wireless network can
potentially change the network topology by adjusting its transmission
power to control its set of 
neighbors.  
The primary goal of
topology control is to design power-efficient algorithms that maintain
network connectivity and optimize performance metrics such as network
lifetime and throughput.   
As pointed out by 
Chandrakasan et.~al \cite{Chandrakasan99},
network protocols that minimize energy consumption are key
to 
the successful usage of 
wireless sensor networks. 
To simplify deployment and reconfiguration
upon failures and mobility, distributed topology control algorithms
that utilize only local information and allow asynchronous operations
are particularly attractive.

The topology control problem can be formalized as follows: We are
given a set $V$ of possibly mobile nodes located in the 
plane.  Each node $u\in V$ is specified by its coordinates,
$(x(u),y(u))$ at any given point in time. Each node $u$ has a power
function $p$ where $p(d)$ gives the minimum power needed to establish
a communication link to a node $v$ at distance $d$ away from $u$.
Assume that the maximum transmission power $P$ is the same for every node,
and the maximum distance for any two nodes to communicate directly is $R$,
i.e. $p(R)=P$.
If every node transmits with power $P$, then we have an
induced graph $G_R = (V,E)$ where $E = \{ (u,v) |\, d(u,v) \leq R\}$
(where $d(u,v)$ is the Euclidean distance between $u$ and $v$).

It is undesirable to have nodes transmit with maximum power for two
reasons.
First, since the power required to transmit between nodes increases
as the $n$th power of the distance between them, for some $n \geq 2$
\cite{Rap96}, it 
may require less power for a node $u$ to relay messages through a series
of 
intermediate nodes to $v$ than to transmit directly to $v$.
In addition, 
the greater the power with which a node transmits, the greater the
likelihood of the transmission interfering with other transmissions.

Our goal in performing topology control is to find a subgraph
$G$ of 
$G_R$ 
such that (1) $G$ consists of all the nodes in $G_R$ but has
fewer edges, (2) if $u$ and $v$ are connected in $G_R$, they are still
connected in $G$, and (3) a node $u$ can transmit to all its neighbors
in $G$ using less power than is required to transmit to all its
neighbors in $G_R$. 
Since minimizing power consumption is so important, it is desirable to
find a graph $G$ satisfying these three properties that minimizes the
amount of power that a node needs to use to communicate with all its
neighbors.
%
For a topology control algorithm to be useful in practice, it must 
be possible for each node
$u$ in the network to construct its neighbor set $N(u)=\{v| (u,v) \in G\}$ 
in a distributed fashion.
Finally, 
if $G_R$ changes to $G_R'$ due to node failures or mobility, 
it  
must be possible to reconstruct a connected $G'$ without global coordination.

In this paper we consider a cone-based topology-control algorithm
introduced in \cite{ourInfocom01}, and show that it satisfies all these
desiderata.  
Most previous papers on topology control have 
utilized position information, which usually requires the availability 
of GPS at each node.
There are a number of disadvantages with using GPS.  In particular, the
acquisition of GPS location information incurs a high delay, and GPS
does not work in indoor environments or cities.
By way of contrast, the cone-based algorithm requires
only the availability of directional information.  That is, it must be
possible to estimate the direction from which another node is
transmitting. 
Techniques for estimating direction 
without requiring position
information are available, and discussed in the IEEE antenna and
propagation  community as the Angle-of-Arrival problem.  
The standard way of doing this is by
using more than one directional
antenna (see \cite{Krizman97}).%
\footnote{Of course, if GPS information is available, 
a node can simply piggyback its location to its message and the required
directional information can be calculated from that.}
\commentout{ 
 The {\emph cone-based} algorithm \cite{ourInfocom01}
takes as a parameter an angle $\alpha$.  A node $u$ finds the minimum
power $p_{u,\alpha}$ such that transmitting with $p_{u,\alpha}$
ensures that in every cone (angular sector) with angle $\alpha$
anchored at $u$ there is a neighbor node.  It is shown that $\alpha
\leq 2\pi/3$ is sufficient to preserve network connectivity.    
} 

The cone-based algorithm takes as a parameter an angle $\alpha$.  A node $u$ then
tries to find the minimum power $p_{u,\alpha}$ such that transmitting
with $p_{u,\alpha}$ ensures that in every cone of degree $\alpha$
around $u$, there is some node that $u$ can reach with power
$p_{u,\alpha}$.  In \cite{ourInfocom01}, it is shown that taking
$\alpha \leq 2\pi/3$ is sufficient to preserve network connectivity.
That is, let $G_{\alpha}$ be the symmetric closure of the
communication graph that results when every node transmits with power
$p_{u,\alpha}$ (so that the neighbors of $u$ in $G_\alpha$ are exactly
those nodes that $u$ can reach when transmitting with power
$p_{u,\alpha}$ together with those nodes $v$ that can reach $u$ by
transmitting with power $p_{v,\alpha}$).  Then it is shown that if
there is a path from $u$ to $v$ in  $G_R$, then there is also such a
path in $G_{2\pi/3}$.   
Moreover, it is also shown that for a reasonable class
of power cost functions and for $\alpha \leq \pi/2$, the network has
competitive power consumption.
More precisely, given arbitrary nodes $u$ and $v$, it is shown that the
power used in the most power-efficient route between $u$
and $v$ in $G_\alpha$ is no worse than $k + 2k \sin(\alpha/2)$ times the
power used in the most power-efficient route in $G_R$ 
(where $k$ is a constant that depends on the 
power consumption model;
if only transmission power is considered and the transmission power $p(d)$
is proportional to the $n$th power of the distance $d$, we
have $k=1$). 
Finally, some optimizations to the basic algorithm are presented.
In the present paper, we
show that taking $\alpha = 5\pi/6$ is necessary and sufficient to
preserve connectivity.  That is, we show that if $\alpha \le 5\pi/6$,
then there is a path from $u$ to $v$ in $G_R$ iff there is such a path
in $G_{\alpha}$  (for all possible node locations) and that if $\alpha
> 5\pi/6$, then there exists a graph $G_R$ that is connected while
$G_\alpha$ is not.  Moreover, we propose new optimizations and
show that they preserve connectivity.  Finally, we discuss how the
algorithm can be extended to deal with dynamic reconfiguration and
asynchronous operations.

There are a number of other papers in the literature on topology
control; as we said earlier, all assume that position information is
available.  Hu \cite{Hu93} describes an
algorithm that does topology control 
using heuristics based on a Delauney triangulation of the graph.  
There 
seems to be 
no guarantee that the heuristics preserve connectivity.
Ramanathan and 
Rosales-Hain \cite{Ramanathan00} 
describe
a centralized spanning tree
algorithm for achieving connected and biconnected static networks,
while minimizing the maximum transmission power. (They also describe 
distributed algorithms that are based on heuristics and
are not guaranteed to preserve connectivity.)
Rodoplu and Meng \cite{Rodoplu99} 
propose
a distributed
position-based topology 
control algorithm that preserves connectivity; 
their algorithm is improved by Li and Halpern \cite{ourICC01}.
In a different vein is the  work described in \cite{peleg00,kg92};  
although it does not deal directly with topology control,
the notion of $\theta$-graph used in these papers bears some resemblance
to the 
cone-based idea described in this paper.
%
Relative neighborhood graphs \cite{tou80} and their relatives
(such  as Gabriel graphs, or $G_{\beta}$ graphs \cite{jt92})
are similar in spirit to the graphs produced by the cone-based
algorithm. 

The rest of the paper is organized as follows.  
Section~\ref{sec-cone} presents
the basic cone-based algorithm and 
shows that $\alpha=5 \pi / 6$ is necessary and
sufficient for connectivity.
Section~\ref{sec-optimizations}
describes several optimizations to the basic algorithm and proves
their correctness.
Section~\ref{sec-reconfig} extends the 
basic algorithm so that it can handle the reconfiguration necessary to
deal with failures and mobility.
Section~\ref{sec-experiments} briefly describes some network simulation 
results 
that show the effectiveness of the basic approach and the optimizations.
Section~\ref{sec-conclusion} concludes the paper.

\section{The Basic Cone-Based Topology Control (CBTC) Algorithm}
\label{sec-cone}

We consider three communication primitives: broadcast, send, and receive,
defined as follows:
\begin{itemize}
\item $\bcast(u,p,m)$ 
is invoked by node $u$ to send message $m$ with power $p$; 
it results in all nodes in the set $\{v|p(d(u,v)) \leq p\}$  receiving $m$.
\item $\send(u,p,m,v)$ is invoked by node $u$ to sent message
$m$ to $v$ with power $p$.
This primitive is used to send unicast messages, i.e. point-to-point
messages.
\item $\recv(u,m,v)$ is used by $u$ to receive message $m$ from $v$. 
\end{itemize}

We assume that when $v$ receives a message $m$ from $u$, it
knows the reception power $p'$ of message $m$.  
This is, in general, less than the power $p$ with
which $u$ sent the message, because of radio signal attenuation in
space.  
Moreover, we assume that, given the transmission power $p$ and the
reception power $p'$, $u$ can estimate $p(d(u,v))$.
This assumption is reasonable in practice.

For ease of presentation, we first assume a synchronous model; that is,
we assume that
communication proceeds in rounds, governed by a global clock, with
each round taking one time unit. 
(We deal with asynchrony in 
Section~\ref{sec-reconfig}.)
In each round each node $u$ can
examine the messages sent to it,
compute, and send messages using the $\bcast$ and $\send$ communication 
primitives.
The communication channel is reliable. 
We later
relax this assumption, and show that the algorithm is correct even in an
asynchronous setting.


The basic Cone-Based Topology Control (CBTC)
algorithm is easy to explain.  The algorithm takes as a parameter an
angle $\alpha$.
Each node $u$ tries 
to find at
least one  neighbor in every cone of degree $\alpha$ centered at $u$.
Node $u$ starts running the algorithm by 
broadcasting a ``Hello'' message using low transmission power, and 
collecting replies.  It gradually
increases the transmission power to discover more neighbors.
It keeps a list of the nodes that it has discovered and the direction in
which they are located.  (As we said in the introduction, we assume that
each node can estimate directional information.)  It then checks whether
each cone of degree $\alpha$ contains a node.  This check is easily
performed: the nodes are sorted according to their angles relative to 
some reference node (say, the first node from which $u$ 
received a reply).  
It is immediate that there is a gap of more than $\alpha$
between the angles of two consecutive nodes iff there is a cone of
degree $\alpha$ centered at $u$ which contains no nodes.  If there is
such a gap, then $u$ broadcasts with greater power.  This continues
until either $u$ finds no $\alpha$-gap or $u$ broadcasts with maximum
power.

Figure~\ref{fig-CBTC} gives the basic CBTC algorithm.  In the algorithm,
a ``Hello'' message is originally broadcasted using some minimal power
$p_0$.  
In addition, the power used to broadcast the message is included in the
message. 
The power is then increased at each step using some function
$\Increase$. 
As in \cite{ourICC01} (where a similar function is used, in the context of a
different algorithm), in this paper, we do not investigate how to choose
the initial power $p_0$, nor do we investigate how to increase the power
at each step.  We simply assume some function $\Increase$ such that
$\Increase^k(p_0) = P$ for sufficiently large $k$.  As observed in
\cite{ourICC01}, an obvious choice is to 
take $\Increase(p) = 2p$.  If the initial choice of $p_0$ is less than
the total power actually needed, then it is easy to see that this
guarantees that $u$'s estimate of the transmission power needed to reach
a node $v$ will 
be within a factor of 2 
of the minimum transmission power actually needed to reach $v$.
Upon receiving a ``Hello'' message from $u$,
node $v$ responds with an {\sl Ack} message.
(Recall that we have assumed that $v$ can compute the power required to
respond.)  
Upon receiving the {\sl Ack} from $v$, node $u$ adds $v$ to its set
$N_u$ of neighbors and adds $v$'s direction $\dir_u(v)$ (measured as an angle
relative to some fixed angle) to its set $D_u$ of directions.
(Recall that we have assumed that $u$ can compute this angle.)
The test $\gap$-$\alpha(D_u)$ tests if there is a gap greater than
$\alpha$ in the angles in $D_u$.

%

\begin{figure}[ht]
\begin{tabbing}
\sc{CBTC($\alpha$)}   \\
\\
$N_u \gets \emptyset$; //the set of discovered neighbors of $u$ \\
$D_u \gets \emptyset$; //the directions from which 
the Acks have come\\
$p_u \gets p_0$; \\
\\
{\bf while}  \=($p_u < P$ and $\gap$-$\alpha(D_u)$) {\bf do} \\
\>    $p_u \gets Increase(p_u)$;\\
\>    $\bcast(u,p_u,($``Hello'',$p_u))$ and gather Acks; \\
\>   $N_u$ $\gets$ $N_u \cup \{v: v$ discovered$\}$;    \\
\>   $D_u \gets D_u \cup \{\dir_u(v): v$ discovered$\}$
\end{tabbing}
\caption{The basic cone-based algorithm running at each node $u$.
\label{fig-CBTC}
}
\end{figure}


Let $N_\alpha(u)$ be the final set of discovered neighbors computed by node $u$
at the end of running CBTC($\alpha$); let $p_{u,\alpha}$ be the corresponding final power.
Let $N_\alpha = \{(u,v) \in V \times V: v \in N_\alpha(u)\}$.
Note that the $N_\alpha$ relation
is not symmetric.  As the following example shows, it is possible that
$(v,u)  \in N_\alpha$ but $(u,v) \notin N_\alpha$.

\begin{figure}[ht]
\setlength\tabcolsep{0.1pt}
\begin{center}
\begin{tabular}{c}
\epsfysize=4.5cm \epsffile{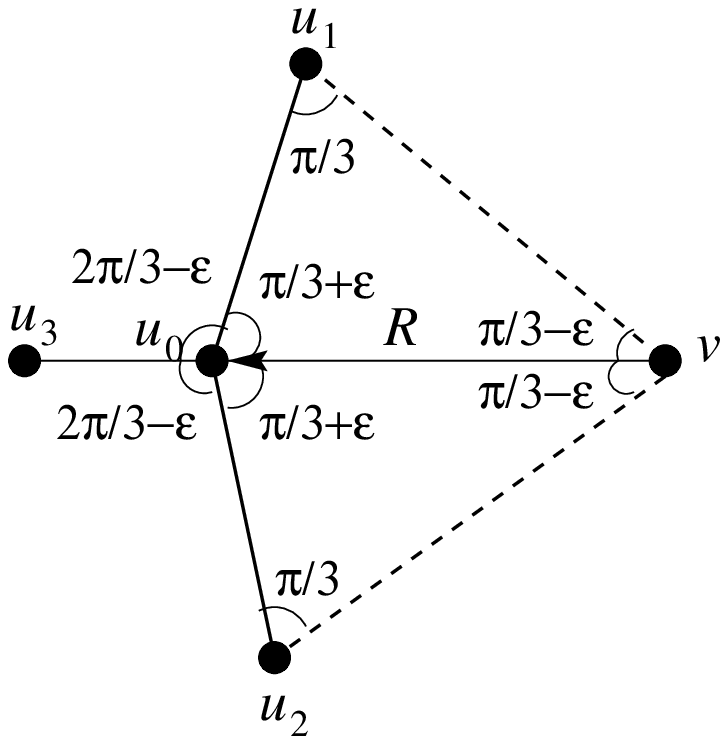}
\end{tabular}
\end{center}
\caption{
$N_\alpha$ may not be symmetric.
\label{fig-counter}
}
\end{figure}

\begin{example}\label{counter} Suppose that $V =
\{u_0,u_1,u_2,u_3,v\}$. (See Figure~\ref{fig-counter}.) 
Further suppose that $d(u_0,v) = R$. 
Choose $\epsilon$ with 
$0 < \epsilon < \pi/12$ 
and place $u_1,
u_2, u_3$ so 
that (1) $\angle{vu_0 u_1} = \angle{vu_0 u_2} =
\pi/3 + \epsilon = \alpha/2$,
(2) $\angle{u_1 v u_0} = \angle{u_2 v u_0} = \pi/3 - \epsilon$ 
(so that $\angle{v u_1 u_0} = \angle{v u_2 u_0} = \pi/3$), 
(3)  $\angle{v u_0 u_3} = \pi$ (so that $\angle{u_1 u_0 u_3} =
\angle{u_2 u_0 u_3} = 2\pi/3 - \epsilon$) and (4) $d(u_0,u_3) = R/2$.
Note that, given $\epsilon$ and the positions of
$u_0$ and $v$,  the positions of $u_1$, $u_2$, and $u_3$ are determined.
Since $\angle {u_1 u_0 v} > \angle{u_0 u_1 v} > \angle {u_1 v u_0} $,
it follows that $d(u_1,v) > d(u_0,v) = R > d(u_0,u_1)$; similarly
$d(u_2,v) > R > d(u_0,u_2)$.  
(Here and elsewhere we use the
fact that, in a triangle, larger sides are opposite larger angles.)
It easily follows that
$N_\alpha(u_0) = \{u_1, u_2, u_3\}$ while $N_\alpha (v) = \{u_0\}$, as
long as 
$2\pi/3 < \alpha \le 5\pi/6$.
Thus, $(v,u_0) \in N_\alpha$, but $(u_0,v) \notin N_\alpha$.
\end{example}
\input{epsf}
%

Let $G_\alpha = (V,E_\alpha)$, where $V$ consists of all nodes in
the network
and $E_\alpha$ is the symmetric closure of $N_\alpha$; that is, 
$(u,v) \in E_\alpha$ iff either $(u,v) \in N_\alpha$ or $(v,u) \in
N_\alpha$.  
We now prove the two main results of this paper: (1) 
if $\alpha \le  
5\pi/6$, then $G_\alpha$ preserves the connectivity of $G_R$ and (2) if
$\alpha > 5\pi/6$, then $G_\alpha$ may not preserve the connectivity of
$G_R$.  Note that Example~\ref{counter} shows the  need for taking the
symmetric closure in computing $G_\alpha$. 
Although $(u_0,v) \in G_R$,
there would be no path from $u_0$ to $v$ if we considered just the edges
determined by $N_\alpha$, without taking the symmetric closure.  
(The fact that $\alpha > 2\pi/3$ in this example is necessary.
As we shall see in Section~\ref{removal}, taking the symmetric closure
is not necessary if $\alpha \le 2\pi/3$.)
As we
have already observed, each node $u$ knows the power required to reach
all nodes $v$ such that $(u,v) \in E_\alpha$: it is just the max of
$p_{u,\alpha}$
and the power required by $u$ to reach each of the nodes $v$ from which
it received a ``Hello'' message.  (As we said earlier, if $u$ receives a
``Hello'' message from $v$, since it 
includes the
power used to transmit it, $u$ can determine the power required for $u$
to reach $v$.)


\begin{theorem}\label{theorem150} If $\alpha \le 5\pi/6$, then $G_\alpha$
preserves the 
connectivity of $G_R$; $u$ and $v$ are connected in $G_\alpha$ iff they
are connected in $G_R$. \end{theorem}

\begin{proof} Since $G_\alpha$ is a subgraph of $G_R$, it is clear that if $u$
and $v$ are connected in $G_\alpha$, they must be connected in $G_R$.
To prove the converse, we start with the following key lemma.

\begin{lemma}
\label{lemma150}
If $\alpha \le 5\pi/6$, and $u$ and $v$ are nodes in $V$ such that
$(u,v) \in E$ (that is, $(u,v)$ is an edge in the graph $G_R$, so that 
$d(u,v) \le R$), then either $(u,v) \in E_\alpha$ or there exist $u', v'
\in V$ such that (a) $d(u',v') < d(u,v)$, (b) either $u' = u$ or $(u,u')
\in E_\alpha$, and (c) either $v' = v$ or $(v,v') \in E_\alpha$.
\end{lemma}

\begin{proof}
A few definitions will be helpful in this and the following proof.
Given two nodes $u'$ and $v'$, 
\begin{itemize}
\item Let $\cone(u',\alpha,v')$ be the cone of
degree $\alpha$ which is bisected by the line $\overline{u'v'}$, as in
Figure~\ref{fig-angle}; 
\item Let $\cir(u,r)$ be the circle centered at $u$ with radius $r$; 
\item Let $\rad_{u,\alpha}^-$ be the distance $d(u,v)$ of the neighbor
  $v$ farthest from $u$ in $N_\alpha(u)$; 
that is, $p(\rad_{u,\alpha}^-)=p_{u,\alpha}$;
\item Let $\rad_{u,\alpha}$ be the distance $d(u,v)$ of the neighbor $v$
  farthest from $u$ in $E_\alpha$.
\end{itemize}
\input{epsf}
\begin{figure}[ht]
\setlength\tabcolsep{0.1pt}
\begin{center}
\begin{tabular}{c}
\epsfysize=4.5cm \epsffile{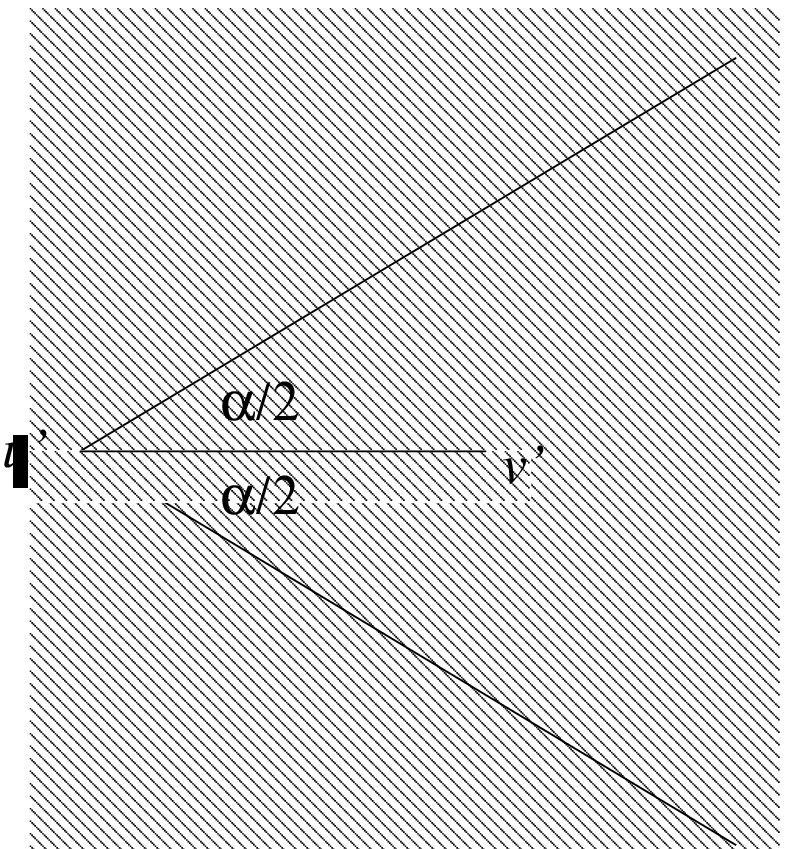}
\end{tabular}
\end{center}
\caption{
$\cone(u',\alpha, v')$
\label{fig-angle}
}
\end{figure}
If $(u,v) \in E_\alpha$, we are done.  Otherwise, it must be the case that
$d(u,v) > \max(\rad_{u,\alpha}^-,\rad_{v,\alpha}^-)$.  
Thus, both $u$ and $v$ terminate CBTC($\alpha$) with no $\alpha$-gap.
It follows that 
$\cone(u,\alpha,v) \inter
N_\alpha(u) \ne \emptyset$ and 
$\cone(v,\alpha,u) \inter N_\alpha(v) \ne \emptyset$. 
Choose $z \in \cone(v,\alpha,u) \inter N_\alpha(v)$ such
that $\angle{zvu}$ is minimal. (See Figure~\ref{fig150}.) Suppose without loss of generality that
$z$ is in the halfplane above $\overline{uv}$.  
If $z$ is actually in $\cone(v,2\pi/3,u)$,
since 
$d(v,z) \le \rad_{v,\alpha}^- < d(u,v)$, 
it follows that 
$d(z,u) < d(u,v)$.  For otherwise, the side $zu$ would be at least as long
as any other side in the triangle $vzu$, so that $\angle{zvu}$ would
have to be at least as large as any other angle in the triangle.  
But since $\angle{zvu} \le \pi/3$, this is impossible.  Thus, taking 
$u' = u$ and $v' = z$, the lemma holds in this case.
So we can assume without loss of generality that $z \notin
\cone(v,2\pi/3,u)$ (and, thus, that 
$\cone(v,2\pi/3,u) \inter N_\alpha(v)
= \emptyset$).  
Let $y$ be the first node in
$N_\alpha(v)$ that a ray that starts at $vz$ would hit 
as it sweeps past $vu$ going counterclockwise.
By construction, $y$ is in the half-plane below $\overline{uv}$ and
$\angle{zvy} \le \alpha$.

Similar considerations show that, without loss of generality, we can
assume that 
$\cone(u,2\pi/3,v) \inter N_\alpha(u) = \emptyset$, and that
there exist two points $w, x \in N_\alpha(u)$ such that (a) $w$ is in
the halfplane above $\overline{uv}$, (b) $x$ is in the halfplane below
$\overline{uv}$, (c) at least one of $w$ and $x$ is in $\cone(u,\alpha,v)$,
and (d) $\angle{wux} \le \alpha$. 
See Figure~\ref{fig150}.

\commentout{
Let $x$ be the first node if $N_\alpha(v)$ that a ray that starts at 
$uw$ would hit as it sweeps past $uv$.  CBTC($\alpha$) guarantees that
$\angle{wux} \le \alpha$.  Similarly, let $y$ be the first node in
$N_\alpha(u)$ that a ray that starts at $vz$ would hit as it sweeps past
$vu$; again, $\angle{zuy} \le \alpha$.  See Figure~\ref{fig150}.
relative to $\overline{uv}$.  
} 

\input{epsf}
\begin{figure}[ht]
\setlength\tabcolsep{0.1pt}
\begin{center}
\begin{tabular}{c}
\epsfysize=6.5cm \epsffile{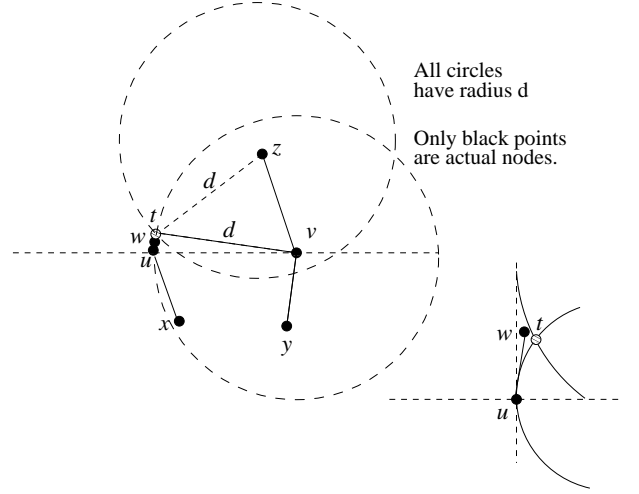}
\end{tabular}
\end{center}
\caption{ Illustration for the proof of Lemma~\ref{lemma150}.
\label{fig150}
}
\end{figure}

If $d(w,v) < d(u,v)$, then the lemma holds with $u' = w$ and $v' = v$,
so we can assume that $d(w,v) \ge d(u,v)$.  Similarly, we can assume
without loss of generality that $d(z,u) \ge d$.
We now prove that $d(w,z)$ and
$d(x,y)$ cannot  both be greater than or equal to $d$.  
This will complete the proof since, for example, if $d(w,z) < d$, then
we can take $u' = w$ and $v' = z$ in the lemma.

Suppose, by way of contradiction, that
$d(w,z)\geq d$ and $d(x,y) \geq d$.  Let $t$ be the
intersection point of $\cir(z,d)$ and $\cir(v,d)$ 
that is closest to $u$.
Recall that at least one of $w$ and $x$ is in $\cone(u,\alpha,v)$.  
As we show in the full paper,
since node $w$ must be
outside (or on) both circles 
$\cir(z,d)$ and $\cir(v,d)$, we have
$\angle{wuv} \geq \angle{tuv}$ (see the closeup on the far
right side of 
Figure~\ref{fig150}).
  
Since $d(t,z) = d(t,v) = d(u,v) = d$, and $d(z,v) < d$, it follows that
$\angle{zvt} > \pi/3$.  Thus,
$$\begin{array}{c}
\angle{tvu} = \angle{zvu} - \angle{zvt} < \angle{zvu} - \pi/3 \mbox{ and} \\
\angle{tvu} = \pi - 2 \times \angle{tuv},
\end{array}$$
and so 
$$\begin{array}{c}
\angle{zvu} - \pi/3 > \pi - 2 \times \angle{tuv} \mbox{ and},\\
\angle{tuv} > 2\pi/3 - \angle{zvu} / 2.
\end{array}$$
Since 
$\angle{wuv} \ge \angle{tuv}$, 
we have that
\begin{equation}
\label{eq-apx4}
\angle{wuv} > 2\pi/3  - \angle{zvu} / 2.
\end{equation}
By definition of $z$, $\angle{zvu} \leq \alpha/2 \leq 5\pi/12$, 
so $\angle{wuv} > 2\pi/3 - 5\pi/24 = 11\pi/24 > \alpha/2$.  
Thus, it must be the case that $w \notin cone(u,\alpha,v)$, so 
$x \in cone(u,\alpha,v)$.

Argument identical to those used to derive (\ref{eq-apx4}) (replacing
the role of $w$ and $z$ by 
$y$ and $x$, respectively) can be used to
show that
%
\begin{equation}
\label{eq-apx5}
\angle{yvu} > 2\pi/3 - \angle{xuv} / 2
\end{equation}
From (\ref{eq-apx4}) and (\ref{eq-apx5}), we have
$$\begin{array}{ll}
&\angle{wuv}  + \angle{xuv} \\
> & (2\pi/3 - \angle{zvu} / 2) +  (4\pi/3 - 2 \times \angle{yvu}) \\ 
= & 2\pi- \angle{zvu} / 2   - 2 \times \angle{yvu}
\end{array}$$
Since $\angle{wuv}$  $+$ $\angle{xuv}$ $\leq$ $\alpha$ $\leq$ $5\pi/6$,
we have that
$5\pi/6$ $>$ $2\pi -
\angle{zvu} / 2$ $-$ $2 \times \angle{yvu}.$ 
Thus, 
$$\angle{zvu} / 2   + 2 \times \angle{yvu} = ((\angle{zvu} +
\angle{yvu}) + 3\times \angle{yvu})/2 > 7\pi/6.$$
Since  $\angle{zvu}$  $+$ $\angle{yvu}$ $\leq$ $\alpha$ $\leq$
$5\pi/6$, it easily follows that $\angle{yvu} > \pi/2$. 
%
As we showed earlier, $\angle{zvu} \ge \angle{zvt} > \pi/3$. 
Therefore, $\angle{zvu}$  $+$ $\angle{yvu}$ $>$ $5\pi/6$. 
This is a contradiction.
\end{proof}

\commentout{
\begin{lemma}
\label{lemma150.1}
If $\alpha \leq 5\pi/6$, and $u$ and $v$ are nodes in $V$ such that 
$(u,v) \in E$, then either $(u,v) \in E_\alpha$ or there exists a path
$u_0 \ldots u_k$ in $G_\alpha$ such that $u_0 = u$, $u_k = v$, $(u_i,
u_{i+1}) \in E_\alpha$, and $d(u_i,u_{i+1}) < d(u,v)$, for $i = 0,
\ldots, k-1$.
\end{lemma}

\begin{proof}
}

The proof of Theorem~\ref{theorem150} now follows easily.
Order the edges in $E$ by length.  We proceed by induction on the 
the rank of the edge in the ordering, using Lemma~\ref{lemma150}, to
show that if $(u,v) \in E$, then there is a path from $u$ to $v$ in
$G_\alpha$.   It immediately follows that if $u$ and $v$ are connected
in $G_R$, then there is a path from $u$ to $v$ in $G_\alpha$.
\qed
\end{proof}

The proof of Theorem~\ref{theorem150} gives some extra information, which
we cull out as a separate corollary:
\begin{corollary}\label{cor150}
If $\alpha \leq 5\pi/6$, and $u$ and $v$ are nodes in $V$ such that 
$(u,v) \in E$, then either $(u,v) \in E_\alpha$ or there exists a path
$u_0 \ldots u_k$ such that $u_0 = u$, $u_k = v$, 
$(u_i,u_{i+1}) \in E_\alpha$, 
and $d(u_i,u_{i+1}) < d(u,v)$, for $i = 0,
\ldots, k-1$.
\end{corollary}

\commentout{
\begin{theorem}
\label{theorem150}
Cone-based algorithm with degree $\alpha$, $\alpha \leq 5\pi/6$,
guarantees
a connected graph;
that is, graph $G_r$ is connected.
\end{theorem}

\begin{proof}
Given any edge $(u,v)$ in $G_R$, we must have $d(u,v) \leq R$.  If the
edge does not exist in $G_r$, $u$ and $v$ must be connected by a path
according to Lemma~\ref{lemma150.1}.  Since $G_R$ is connected, $G_r$
must be connected.
\end{proof}
} 


Next we prove that degree $5\pi/6$ is a tight upper bound; if 
$\alpha > 5\pi/6$, then CBTC($\alpha$) does not necessarily preserve
connectivity.

\begin{theorem}
\label{theorem150-counter}
If $\alpha > 5\pi/6$, then CBTC($\alpha$) does not necessarily preserve
connectivity. 
\end{theorem}

\input{epsf}
\begin{figure}[ht]
\setlength\tabcolsep{0.1pt}
\begin{center}
\begin{tabular}{c}
\epsfysize=6.5cm \epsffile{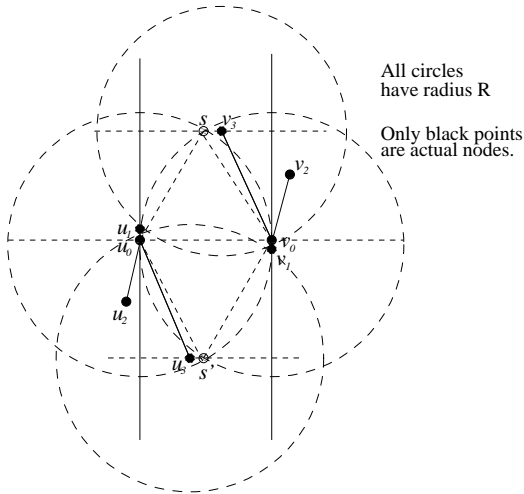}
\end{tabular}
\end{center}
\caption{
A disconnected  graph if  $\alpha = 5\pi/6+\epsilon$.
\label{fig-counter150}
}
\end{figure}

\begin{proof}
Suppose $\alpha = 5\pi/6+\epsilon$ for some $\epsilon > 0$. 
We construct a graph $G_R = (V,E)$ such that CBTC($\alpha$) does not preserve
the connectivity of this graph.  $V$ has eight nodes: 
$u_0, u_1, u_2, u_3, v_0, v_1, v_2, v_3$. (See Figure~\ref{fig-counter150}.) We call $u_0, u_1, u_2, u_3$ the
{\em $u$-cluster}, and $v_0, v_1, v_2, v_3$ the {\em $v$-cluster}.  The
construction has the property that 
$d(u_0,v_0) = R$ and for $i,j  = 0, 1, 2, 3$, we have $d(u_0,u_i) < R$,
$d(v_0,v_i) < R$, and $d(u_i,v_j) > R$ if $i+j \ge 1$.  That is, the only
edge between the $u$-cluster and the $v$-cluster in $G_R$ is $(u_0,v_0)$.  
However, in $G_{\alpha}$, the $(u_0,v_0)$ edge disappears, so that 
the u-cluster and the v-cluster are disconnected.

In Figure~\ref{fig-counter150}, $s$ and $s'$ are the intersection points of the circles of
radius $R$ centered at $u_0$ and $v_0$, respectively.  Node $u_1$ is
chosen so that $\angle{u_1u_0v_0} = \pi/2$.  Similarly,
$v_1$ is chosen so that
 $\angle{v_1v_0u_0}
= \pi/2$ and $u_1$ and $v_1$ are on opposite sides of the line
$\overline{u_0v_0}$.  Because of the right angle, it is clear that,
whatever $d(u_0,u_1)$ is, we must have
$d(v_0,u_1) > d(v_0,u_0) = R$; similarly, $d(u_0,v_1) > R$ whatever
$d(v_0,v_1)$ is.   Next,
choose $u_2$ so that 
$\angle{u_1u_0u_2} = \min(\alpha,\pi)$ 
and
$u_0u_2$ comes after $u_0u_1$ as a ray sweeps around counterclockwise
from $u_0v_0$.   It is easy to see that $d(v_0,u_2) > R$, whatever
$d(u_0,u_2)$ is, since $\angle{v_0u_0u_2} \ge \pi/2$.  For
definiteness, choose $u_2$ so that $d(u_0,u_2) = R/2$.
Node $v_2$ is
chosen similarly.  The key step in the construction is the choice of
$u_3$ and $v_3$.  Note that $\angle{s'u_0u_1} = 5\pi/6$.   Let 
$u_3$ be a point on the line through $s'$ parallel to
$\overline{u_0v_0}$ slightly to the left of $s'$ such that
$\angle{u_3u_0u_1} < \alpha$.  Since $\alpha = 5\pi/6 + \epsilon$, it is
possible to find such a node $u_3$.  Since $d(u_0,s') = d(v_0,s') = R$
by construction, it follows that $d(u_0,u_3) < R$ and $d(v_0,u_3) > R$.
It is clearly possible to choose $d(v_0,v_1)$ sufficiently small so that 
$d(u_3,v_1) > R$.  The choice of $v_3$ is similar.

It is now easy to check that when $u_0$ runs CBTC($\alpha$), it will
terminate with $p_{u_0,\alpha} = \max(d(u_0,u_3),R/2) < R$; similarly
for $v_0$.  Thus, this construction has all the required properties.
\end{proof}

\section{Optimizations}
\label{sec-optimizations}

In this section, we describe three optimizations to the basic
algorithm. 
We prove that these optimizations allow some
of the edges to be removed 
while still preserving connectivity.

\subsection{The shrink-back operation}\label{shrinkback}
In the basic CBTC($\alpha$) algorithm, $u$ is said to be a 
{\em boundary node\/} if, at the end of the algorithm, $u$ still has an 
$\alpha$-gap.  Note that this means that, at the end of the algorithm, a
boundary node broadcasts with maximum power.  
An optimization,
sketched in \cite{ourInfocom01},
would be to add a shrinking phase at the end of the
growing phase to allow each boundary node to broadcast with less power,
if it can do so 
without reducing its cone coverage.  
To make this precise, given a set $\dir$ of directions (angles) and an
angle $\alpha$, define $\cov_\alpha(\dir) = \{\theta:$ for some
$\theta' \in \dir$, $|\theta - \theta'| 
\bmod 2\pi \le \alpha/2\}$.  
We modify CBTC($\alpha$) so that, at each iteration, a node in $N_u$ is
tagged with the power used the first time it was discovered.  Suppose
that the power levels used by node $u$ during the algorithm were $p_1,
\ldots, p_k$. If $u$ is a boundary node, $p_k$ is the maximum power $P$.  
A boundary node successively removes nodes tagged with power $p_k$, then
$p_{k-1}$, and so on, as long as their removal does not change the
coverage.  
That is, let $\dir_i$,  $i = 1, \ldots, k$, be the set of directions found with all
power levels $p_i$ or less, then the minimum $i$ such that
$\cov_\alpha(\dir_i) = \cov_\alpha(\dir_k)$ is found. 
Let $N_\alpha^s(u)$ consist of all the nodes in $N_\alpha(u)$ tagged
with power $p_i$ or less.  Let $N_\alpha^s = \{(u,v): v \in
N_\alpha^s(u)\}$, and let $E_\alpha^s$ be the symmetric closure of
$N_\alpha^s$.  Finally, let $G^s_\alpha = (V,E_\alpha^s)$.



\begin{theorem}
\label{theorem-shrink}
If  $\alpha \leq 5\pi/6$, 
then $G_\alpha^s$ preserves the connectivity of $G_R$.
\end{theorem}

\begin{proof}
It is easy to check that the proof of Theorem~\ref{theorem150} depended
only on the cone coverage of each node, so it goes through without change.
\end{proof}

Note that this argument actually shows that we can remove any nodes from
$N_u$ that do not contribute to the cone coverage.  
However, our interest here lies in minimizing power, not in
minimizing the number of nodes in $N_u$.
There may be some applications where it helps to reduce the 
degree of
a node; in this case, removing further nodes may be a useful
optimization.  


\subsection
{Asymmetric  
edge removal}\label{removal}

As shown by Example~\ref{counter}, in order to preserve connectivity, it
is necessary to add an edge $(u,v)$ to $E_\alpha$ if $(v,u) \in
N_\alpha$, even if $(u,v) \notin N_\alpha$.  
In Example~\ref{counter}, $\alpha > 2\pi/3$.  
This is not an accident.  As we now show,
if $\alpha \le 2\pi/3$, not only don't we have to add an edge $(u,v)$ if
$(v,u) \in N_\alpha$, we can {\em remove\/} an edge $(v,u)$ if $(v,u) \in
N_\alpha$ but $(u,v) \notin N_\alpha$.  Let $E_\alpha^- = \{(u,v): (u,v)
\in N_\alpha$ and $(v,u) \in N_\alpha\}$.  Thus, while $E_\alpha$ is the
smallest symmetric set containing $N_\alpha$, $E_\alpha^-$ is the
largest symmetric set contained in $N_\alpha$.  Let $G_\alpha^- =
(V,E_\alpha^-)$. 

\begin{theorem}\label{theorem120} If $\alpha \le 2\pi/3$, then 
$G_\alpha^-$
preserves the connectivity of $G_R$.
\end{theorem}

\begin{proof} 
We start by proving the following lemma, which 
strengthens Corollary~\ref{cor150}.

\begin{lemma}
\label{lemma120}
If $\alpha \leq 2\pi/3$, and $u$ and $v$ are nodes in $V$ such that 
$(u,v) \in E$, then 
either $(u,v) \in N_\alpha$ or
there exists a path
$u_0 \ldots u_k$ such that $u_0 = u$, $u_k = v$, $(u_i,
u_{i+1}) \in N_\alpha$, and $d(u_i,u_{i+1}) < d(u,v)$, for $i = 0,
\ldots, k-1$.
\end{lemma}

\begin{proof}
Order the edges in $E$ by length.  We proceed by strong induction on the rank
of an edge in the ordering.  Given an edge $(u,v) \in E$ of rank $k$ in
the ordering, if $(u,v) \in N_\alpha$, we are done.  If not, 
as argued in the proof of Lemma~\ref{lemma150}, there must be
a node $w \in \cone(u,\alpha,v) \inter N_\alpha(u)$.
Since $\alpha \le 2\pi/3$, the argument in the proof of
Lemma~\ref{lemma150} also shows that $d(w,v) < d(u,v)$.  Thus, $(w,v)
\in E$ and has lower rank in the ordering of edges.  Applying the
induction hypothesis, the lemma holds for $(u,v)$.   This completes the
proof.
\end{proof}

Lemma~\ref{lemma120} shows that if $(u,v) \in E$, then there is a path
consisting of edges in $N_\alpha$ from $u$ to $v$.  This is not good
enough for our purposes; we need a path consisting of edges in
$E_\alpha^-$.  The next lemma shows that this is also possible.
\begin{lemma}
\label{lemma120.1}
If $\alpha \leq 2\pi/3$, and $u$ and $v$ are nodes in $V$ such that
$(u,v) \in N_\alpha$, then there exists a path $u_0 \ldots u_k$ such
that $u_0 = u$, $u_k = v$, $(u_i, u_{i+1}) \in E_\alpha^-$, for $i =
0, \ldots, k-1$.
\end{lemma}

\begin{proof}
Order the edges in $N_\alpha$ by length. We
proceed by strong induction on the rank of an edge in the ordering.
Given an edge $(u, v) \in N_\alpha$ of rank $k$ in the ordering, if
$(u, v) \in E^-_\alpha$, we are done.  If not, we must have $(v, u)
\not\in N_\alpha$. Since $(v, u) \in E$, by Lemma~\ref{lemma120}, there is a path
from $v$ to $u$ consisting of edges in $N_\alpha$ all of which have
length smaller than $d(v, u)$.  If any of these edges is in $N_\alpha
- E^-_\alpha$, we can apply the inductive hypothesis to replace  the
edge by a path consisting only of edges in $E^-_\alpha$.  By the
symmetry of $E^-_\alpha$, such a path from $v$ to $u$ implies a path
from $u$ to $v$. This completes the inductive step.
\end{proof}

The proof of Theorem~\ref{theorem120} is now immediate from
Lemmas~\ref{lemma120} and~\ref{lemma120.1}.
\qed
\end{proof}

To implement asymmetric edge removal, the basic CBTC needs to be
enhanced slightly.  
After finishing CBTC($\alpha$), a node $u$ must send a message to 
each node $v$ to which it sent an {\sl Ack} message 
that is not in
$N_\alpha(u)$, telling $v$ to remove $u$ from $N_\alpha(v)$ when
constructing $E_\alpha^-$.  
It is easy to see that the shrink-back optimization discussed in
Section~\ref{shrinkback} can be applied together with the removal of
these asymmetric edges.  

It is clear that there is a tradeoff between using CBTC($5\pi/6$) and
using CBTC($2\pi/3$) with asymmetric edge removal.  In general,
$p_{u,5\pi/6}$ (i.e., $p(\rad_{u,5\pi/6}^-)$) will be smaller than
  $p_{u,2\pi/3}$.  However, 
the power 
$p(rad_{u,5\pi/6})$
with which $u$ needs to transmit may be
greater than $p_{u,5\pi/6}$ since 
$u$ may need to reach nodes $v$ such
that 
$u \in N_{5\pi/6}(v)$ but $v \notin N_{5\pi/6}(u)$.  In contrast,
if $\alpha = 2\pi/3$, 
then asymmetric edge removal allows $u$ to still use $p_{u,2\pi/3}$ and may allow $v$ to use power 
less than $p_{v,2\pi/3}$.
Our experimental results confirm this. See Section~\ref{sec-experiments}.
\subsection{Pairwise edge removal}
%
The final optimization aims at further reducing the transmission power
of each node.
In addition to the directional information, this optimization
requires two other pieces of information. 
First, each node $u$ is assigned a unique integer ID denoted ID$_u$,
and that ID$_u$ is included in all of $u$'s messages. Second,
although a node $u$ does not need to know its exact distance from
its neighbors, given any pair of neighbors $v$ and $w$, node $u$ needs
to know which of them is closer. This can be achieved as
follows. 
Recall that a node grows its radius in discrete steps. It
includes its transmission power level in each of the 
``Hello'' messages.
Each discovered neighbor node also includes its transmission power level in the {\sl Ack}.
When $u$ receives messages
from nodes $v_1$ and $v_2$, it can deduce which of $v_1$ and $v_2$ is closer
based on the transmission and reception powers of the messages.


Even after the shrink-back operation and possibly 
asymmetric
edge removal, there are many edges that can be removed 
while still preserving connectivity.
For example, if three edges form a triangle, we
can clearly remove any one of them while still maintaining connectivity. 
This optimization (where the longest edge is removed) is used
in \cite{ourInfocom01}. In this section, we improve on this result by
showing
that if there is an edge from $u$ to $v_1$ and from $u$ to $v_2$, 
then we can
remove
the longer edge 
even if there is no edge from $v_1$ to $v_2$,
as long as $d(v_1,v_2) < \max(d(u,v_1),d(u,v_2))$.  Note that a
condition sufficient to guarantee that $d(v_1,v_2) <
\max(d(u,v_1),d(u,v_2))$ is that $\angle{v_1uv_2} < \pi/3$ (since the
longest edge will be opposite the largest angle).  

To make this precise, we use the notion of an edge ID.
Each edge $(u,v)$ is assigned an edge ID $eid(u,v)$ $=$ $(i_1$, $i_2$, 
$i_3)$, 
where $i_1$ $=$ $d(u,v)$,  
$i_2$ $=$ $max($ID$_u$, ID$_v)$, and
$i_3$ $=$ $min($ID$_u$, ID$_v$).  
Edge IDs are compared lexicographically, so that $(i,j,k) <
(i',j',k')$ iff either (a) $i < i'$, (b) $i = i'$ and $j < j'$, or (c)
$i = i'$, $j=j'$, and $k < k'$.


\begin{definition}
\label{l-edge}
If $v$ and $w$ are neighbors of $u$, $\angle{vuw} < \pi/3$, and
$eid(u,v) > eid(u,w)$,
then $(u,v)$ is  a {\em redundant edge}. 
\end{definition}
As the name suggests, redundant edges are redundant, in that it is
possible to remove them while still preserving connectivity.
The following theorem proves this.

\begin{theorem}
\label{theorem-longer}
For $\alpha \leq 5\pi/6$, 
all redundant edges
can be removed
while still preserving connectivity.
\end{theorem}

\begin{proof}
Let $E^{nr}_\alpha$ consist of all the non-redundant edges
in $E_\alpha$.
We show that if $(u,v) \in E_\alpha - E^{nr}_\alpha$, then there is a path from
$u$ to $v$ consisting only of edges in $E^{nr}_\alpha$.  Clearly, this
suffices to prove the theorem.

Let $e_1, 
e_2, \cdots, e_m$ be a listing of the redundant edges (i.e, those in
$E_\alpha - E^{nr}_\alpha$) in increasing lexicographic order of edge
ID.  
We prove, by induction on $k$, that for every
redundant edge
$e_k =
(u_k,v_k)$ there is a path from $u_k$ to $v_k$ consisting of
edges in $E^{nr}_\alpha$.
For the base case, consider
$e_1 = (u_1,v_1)$. By definition, there must exist an edge
$(u_1,w_1)$  such that $\angle{v_1u_1w_1} < \pi/3$ and
$eid(u_1,v_1) > eid(u_1,w_1)$.
Since $e_1$ is the
redundant edge
with the smallest edge ID, $(u_1,w_1)$ cannot be a
redundant edge.  Since $\angle{v_1u_1w_1} < \pi/3$, 
it follows that 
$d(w_1,v_1) < d(u_1,v_1)$.
If $(w_1,v_1) \in E_\alpha$,  
then $(w_1,v_1) \in E^{nr}_\alpha$ (since
$(u_1,v_1)$ is the shortest redundant edge) and $(u_1,w_1),(w_1,v_1)$ is
the desired path of non-redundant edges.  On the other hand,
if $(w_1,v_1) \notin E_\alpha$ 
then, since
$d(w_1,v_1) < d(u_1,v_1) \leq R$  and 
$\alpha \leq 5\pi/6$,
by Corollary~\ref{cor150}, 
there exists a path from $w_1$ to $v_1$
consisting of edges in 
$E_\alpha$
all shorter  than 
$d(w_1,v_1)$.
Since none of these edges can be 
redundant edge,
this gives us the desired path.  

For the inductive step,
suppose that for every $e_j =(u_j,v_j)$, $1 \leq j \leq i-1$, we have
found a path $H'_j$ between $u_j$ and $v_j$, which contains no
redundant edges.  
Now consider $e_i  = (u_i,v_i)$.
Again, by definition, there exists another edge $(u_i,w_i)$ with
$eid(u_i,v_i) > eid(u_i,w_i)$ and $\angle{v_iu_iw_i} < \pi/3$.  If
$(u_i,w_i)$ is a
redundant edge, it must be one of $e_j$'s,
where $j \leq i-1$.  Moreover, if the path $H_i$ 
(from Corollary~\ref{cor150})
between $v_i$ and $w_i$ contains a
redundant edge
$e_j$, we
must have $|e_j| < |e_i|$ and so $j \leq i-1$.  By connecting
$(u_i,w_i)$ with $H_i$ and replacing every
redundant edge
$e_j$ on the path with $H'_j$, we obtain a path $H'_i$ between $u_i$
and $v_i$ that contains no
redundant edges.  This completes the proof.
\end{proof}


Although Theorem~\ref{theorem-longer} shows that all redundant edges
can be removed, this doesn't mean that all of them should necessarily be
removed. For example, if we remove some edges, the paths between
nodes become longer, in general.  Since some overhead is added for each
link a message traverses, having fewer edges can affect network
throughput. In addition, if routes are known and many messages are
being sent using point-to-point communication between different
senders and receivers, having fewer edges is more likely to cause
congestion. Since we would like to reduce the transmission power of
each node, we remove only redundant edges with length greater than the longest
non-redundant edges. We call this optimization the {\it pairwise edge
  removal} optimization. 


\section{Dealing with reconfiguration, asynchrony, and failures}
\label{sec-reconfig}
In a multi-hop wireless network, nodes can be mobile. Even if nodes do
not move, nodes may die if they run out of energy. 
In addition, new nodes may be added to the network.
We need a mechanism to detect such changes in the network.
This is 
done by the Neighbor Discovery Protocol (NDP). A
NDP is usually a simple beaconing protocol for each node to tell its 
neighbor that it is still alive.
The beacon includes the sending node's ID and the transmission power of
the beacon.  
A neighbor is considered failed if a pre-defined number of beacons are
not received for a certain time interval $\tau$. A node $v$ is
considered a new neighbor of $u$ if a beacon is received from $v$ and
no beacon was received from $v$ during the previous $\tau$ interval.

The question is what power a node should use for beaconing.
Certainly a node $u$ should broadcast with sufficient power to reach all 
of its neighbors in $E_\alpha$ (or $E_\alpha^-$, if $\alpha \le 2\pi/3$).
As we will show,  
if $u$ uses a beacon with power $p(\rad_{u,\alpha})$ (recall that
$p(\rad_{u,\alpha})$ is the power that $u$ must use to 
reach all its neighbors in $E_\alpha$), then this
is sufficient 
for reconfiguration to work with
the basic cone-based algorithm
(possibly combined with asymmetric edge removal if $\alpha \le
2\pi/3$,
in which case we can use power $p(\rad_{u,\alpha}^-$)).

We define three basic events: 
\begin{itemize}
\item A $\join_u(v)$ event happens when node $u$ detects 
a beacon from node $v$ for the first time;
\item A ${leave}_u(v)$ event happens when node $u$
misses some predetermined number of beacons from node $v$;
\item An $\angleChange_u(v)$ event happens when $u$ detects that $v$'s
angle with respect to $u$ 
has changed.  (Note this could be due to movement by
either $u$ or $v$.)
\end{itemize} 

Our reconfiguration algorithm is very simple.  It is convenient to
assume that each node is tagged with the power used when it was first
discovered, as in the shrink-back operation.  (This is not necessary,
but it minimizes the number of times that CBTC needs to be rerun.)  
\begin{itemize}
\item If a $\leave_u(v)$ event happens, and if there is an
$\alpha$-gap after dropping $\dir_u(v)$ from $D_u$,
node $u$ reruns 
CBTC($\alpha$) (as in Figure~\ref{fig-CBTC}), 
starting with power $p(\rad_{u,\alpha}^-)$
(i.e., taking $p_0 = p(\rad_{u,\alpha}^-)$).
\item If a $\join_u(v)$ event happens, 
$u$ computes $\dir_u(v)$ and the power needed to reach $v$.  
As in the
shrink-back operation, $u$ then removes nodes, 
starting with the farthest neighbor nodes and working back,
as long as their removal does not
change the coverage.
\item 
If an ${\angleChange}_u(v)$ event happens, node
$u$ 
modifies the set $D_u$ of directions appropriately.  If an $\alpha$-gap
is then detected, then CBTC($\alpha$) is rerun, again starting with power
$p(\rad_{u,\alpha}^-)$.  Otherwise, nodes are removed,
as in the shrink-back operation, to see if less power can be used.
\end{itemize}

In general, there may be more than one change event that is detected at a
given time by a node $u$. (For example, if $u$ moves, then there will
be in general several 
$\leave$, $join$ and $\angleChange$ events detected by $u$.)  If
more than one change event is detected by $u$, we perform the changes
suggested above as if the events are observed in some order, as long as
there is no need to rerun CBTC.  If CBTC needs to be rerun, 
it deals with all changes simultaneously.

Intuitively,
this reconfiguration algorithm preserves connectivity.  We
need to be a little careful in making this precise, since if the
topology changes frequently enough, the reconfiguration algorithm may
not ever catch up with the changes, so there may be no point at which
the connectivity of the network is actually preserved.  Thus, what we
want to show is that if the topology ever stabilizes, so that there are
no 
further changes, then the reconfiguration algorithm eventually results 
in a graph that preserves the connectivity of the final network, as
long as there are periodic 
beacons.  
It should be clear that the reconfiguration algorithm
guarantees that each cone of degree $\alpha$ around a node $u$ is
covered (except for boundary nodes), just as the basic algorithm does.
Thus, the proof that the reconfiguration algorithm preserves
connectivity follows immediately from the proof of
Theorem~\ref{theorem150}. 
\commentout{
In particular, although it
may be necessary from time to time to broadcast with maximum power, this
is taken care of automatically by the reconfiguration algorithm.
(Recall that border nodes will broadcast with maximum power.)
For ease of exposition,
we prove the theorem only for the basic algorithm, without any of the
optimizations discussed in Section~\ref{sec-optimizations}.  

\begin{theorem}\label{thm-reconfigOK}
Suppose that there are periodic regular beacons and that, after a
sequence of topology changes, the network stabilizes at some time $t$
so that it is characterized by an induced graph $G_R^t$, after which
there are no further changes.  
Then at some time $t' > t$, the combination of CBTC($\alpha$), for
$\alpha \ge 5\pi/6$, and the reconfiguration algorithm will result in a
graph that preserves the connectivity of $G_R^t$. 
\end{theorem}

\begin{proof}  [[LI, YOU WILL NEED TO PROVE THIS.  THIS IS THE ONLY
SENSE IN WHICH I CAN THINK OF THAT YOU CAN SAY THE RECONFIGURATION
ALGORITHM IS CORRECT.  I SUSPECT THAT THE PROOF IS NOT DIFFICULT, IF YOU
STRUCTURE IT RIGHT.  SOMEHOW YOU WANT TO ARGUE THAT IT'S CORRECT IF
THERE IS ONLY ONE TOPOLOGY CHANGE, AND THEN CONTINUE BY INDUCTION.  IF
YOU DON'T HAVE THE TIME TO WRITE THIS UP CAREFULLY FOR THE PODC VERSION,
YOU CAN LEAVE IT TO THE FULL PAPER.]]
\end{proof}
} 
%

While this reconfiguration algorithm works in combination with the basic
algorithm CBTC($\alpha$) and in combination with the asymmetric edge
removal optimization, we must be careful in combining it with the other
optimizations discussed in Section~\ref{sec-optimizations}.  In
particular, 
we must be very careful about what power a node should use 
for its beacon.
For example, if the shrink-back operation is
performed, using the power to reach all the neighbors in 
$G_\alpha^s$
does not suffice.
For suppose that the network is temporarily partitioned into two
subnetworks $G_1$ and $G_2$; for every pair of nodes $u_1 \in G_1$ and
$u_2 \in 
G_2$, the distance $d(u_1,u_2) > R$.  
Suppose that $u_1$ is a boundary node in $G_1$ and  $u_2$ is a boundary
node in $G_2$, and that, as a result of the shrink-back operation, both
$u_1$ and $u_2$ use power $P' < P$.
Further suppose that later nodes
$u_1$ and $u_2$ move closer together so that $d(u_1,u_2) < R$.  
If $P'$ is not sufficient power for $u_1$ to communicate with $u_2$,
then they will never be aware of each other's presence, since their
beacons will not reach each other, so they will not detect that the
network has become reconnected.
Thus,  network connectivity is {\em not\/} preserved.

This problem can be solved by having the boundary nodes broadcast with
the power computed by the basic CBTC($\alpha$) algorithm, namely $P$ in
this case.  
Similarly, with the pairwise edge removal optimization, it is necessary
for $u$'s beacon to broadcast with 
$p(\rad_{u,\alpha})$,
i.e., the power needed to reach all of $u$'s
neighbors in $E_\alpha$, not just the power needed to reach all of
$u$'s neighbors in $E_\alpha^{nr}$.  It is easy to see that this choice
of beacon power guarantees that the reconfiguration algorithm works.



It is worth noting that a reconfiguration protocol works
perfectly well in an asynchronous setting.  In particular, 
the synchronous model with reliable channels that has been assumed up to
now  can be relaxed to allow 
asynchrony and both communication and node failures.
Now nodes are assumed to communicate asynchronously, messages may get
lost or duplicated, and nodes may fail (although we consider only {\em crash\/}
failures: either a node crashes and stops sending messages, or it
follows its algorithm correctly).  
We assume that messages have unique identifiers and that mechanisms to
discard duplicate messages are present. 
Node failures result in $\leave$ events, as do lost messages.  If node
$u$ gets a message after many messages having been lost, there will be a
$\join$ event corresponding to the earlier $\leave$ event.  



\section{Experimental Results}
\label{sec-experiments}
In order to understand the effectiveness of our algorithm and its
optimizations, we generated 100 random networks, each with 100
nodes. These nodes are randomly placed in a $1500 \times 1500$
rectangular region. Each node has a maximum transmission radius of
$500$.  

\begin{ctable*}
{
\footnotesize
\begin{tabular}{|l|c|c|c|c|c|c|c|c|}
\hline
            & \multicolumn{2}{c}{Basic} \vline &
            \multicolumn{2}{c}{with $op_1$} \vline &
            \multicolumn{1}{c}{with $op_1$ and $op_2$}\vline  &
            \multicolumn{2}{c}{with all $op$} \vline  & 
Max Power  \\
\hline
  Average   &  $\alpha=5\pi/6$ &  $\alpha=2\pi/3$  & $\alpha=5\pi/6$
 &
            $\alpha=2\pi/3$ & $\alpha=2\pi/3$ & $\alpha=5\pi/6$ & $\alpha=2\pi/3$     &      
\\
\cline{2-8}
Node Degree    & 12.3  & 15.4   & 10.3  & 12.8 & 7.0   &
3.6   & 3.6   & 25.6 \\
\cline{1-9}
Average radius & 436.8 & 457.4 & 373.7 & 398.1 & 276.8 &
155.9 & 160.6 & 500  \\ 
\hline
\end{tabular}
\caption{Average degree and radius of the cone-based topology control
  algorithm with different $\alpha$ and optimizations 
($op_1$--shrink-back,
$op_2$--asymmetric edge removal, $op_3$--pairwise edge removal).
} 
\label{table-avg}
}
\end{ctable*}
In Figure~\ref{fig-topo}, the results from one of these random networks
are used to illustrate
how CBTC and the various optimizations improve network topology.
Figure~\ref{fig-topo}(a) shows a topology graph in which no topology control
is employed and every node transmits with maximum power.
Figures~\ref{fig-topo}(b) and (c) show the corresponding graphs produced by
CBTC($2\pi/3$) and CBTC($5\pi/6$), respectively.
From them, 
we can see that
both CBTC($2\pi/3$) and CBTC($5\pi/6$) allow nodes in the
dense areas to automatically reduce their transmission
radius. Figures~\ref{fig-topo}(d) and (e) illustrate the graphs after
the shrink-back operation is performed.
Figure~\ref{fig-topo}(f) shows the graph for $\alpha=2\pi/3$ as a
result of the shrink-back operation and the asymmetric edge removal.
Figures~\ref{fig-topo}(g) and (h) show the topology graphs after
all applicable optimizations.

Table~\ref{table-avg} compares the cone-based algorithm with $\alpha=2\pi/3$
and $\alpha=5\pi/6$ in terms of average node degree and average
radius. 
It also shows the effect of transmitting at maximum power (i.e., with no
attempt at topology control.)
The results are averaged over the 100 random networks mentioned earlier.
As expected, 
using a larger value for $\alpha$ results in a smaller node degree 
and radius.
However, 
as we discussed in Section~\ref{removal}, 
there is a tradeoff between using CBTC($2\pi/3$) and
CBTC($5\pi/6$). 
Just using the basic algorithm results in $\rad_{u,5\pi/6}=436.8 <
\rad_{u,2\pi/3} = 457.4$. 
But after applying asymmetric edge removal with $\alpha = 2\pi/3$, 
the resulting radius is 301.2
(this number is not shown in the table);
asymmetric edge removal can result in significant savings.
After applying all applicable optimizations, both $\alpha = 2\pi/3$ and
$\alpha = 5\pi/6$ end up with essentially the same
average node degree of 3.6 and almost the same average radius.
However, there are some secondary advantages to take $\alpha = 5\pi/6$.  
In general, CBTC($5\pi/6$) will terminate sooner than CBTC($2\pi/3$) and
so expend less power during its execution (since $p_{u,5\pi/6} <
p_{u,2\pi/3}$).  
Thus, especially if
reconfiguration happens often, there are advantages to using CBCT($5\pi/6$).

The last column in Table~\ref{table-avg} gives the performance numbers
for the case of no topology control,
under the assumption that each  node uses
the maximum transmission power of $p(500)$.
Using topology control cuts down the average degree by a factor of more
than 7 (3.6 vs.~25.6) and cuts down the average radius by a factor of
more than 3 (155.9 or 160.6 vs.~500).  Clearly, this is a significant
improvement.  
\input{epsf}
\begin{figure*}[ht]
\begin{center}
\begin{tabular}{cc}
\epsfysize=5.0cm \epsffile{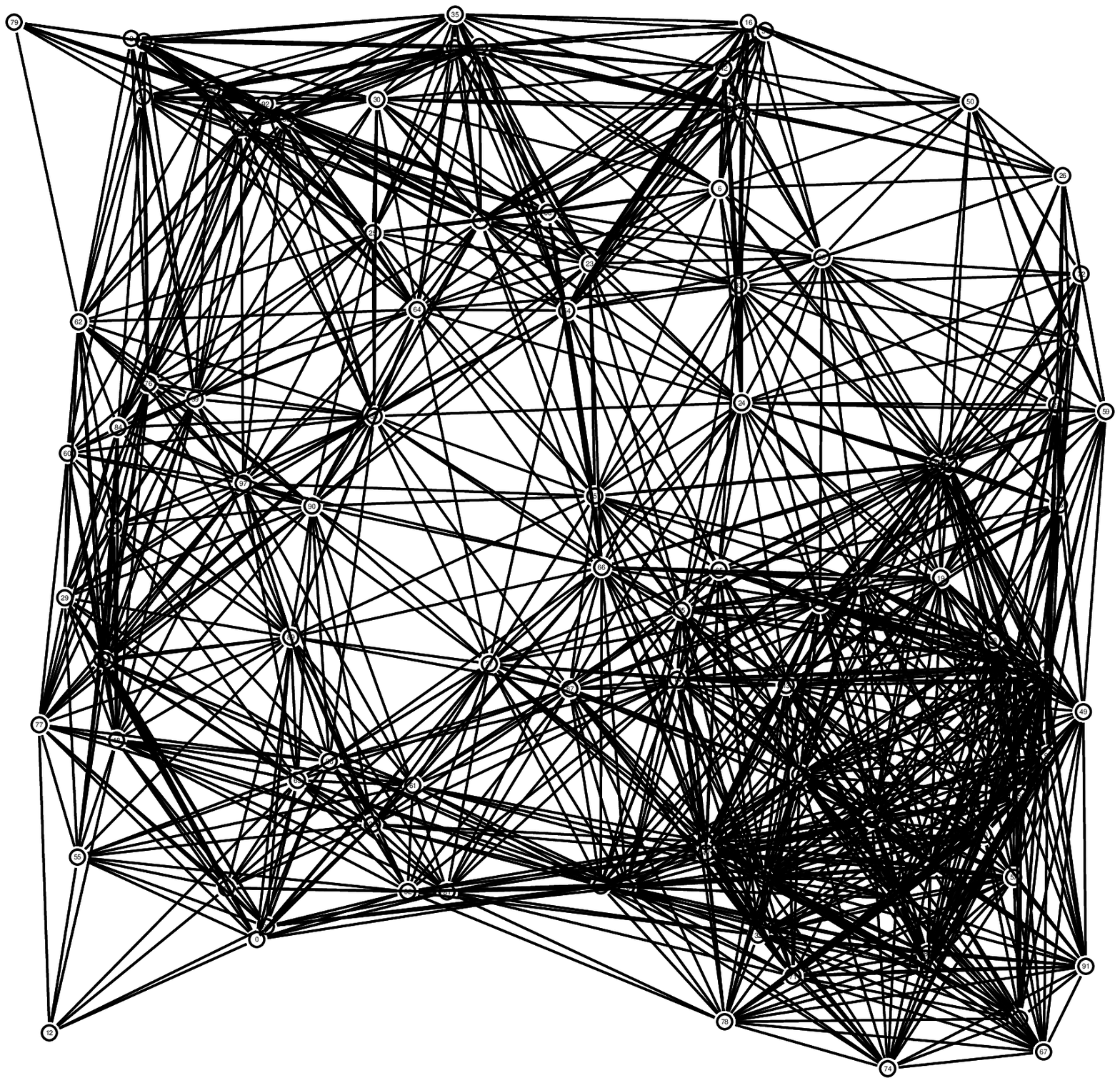} & 
\epsfysize=5.0cm \epsffile{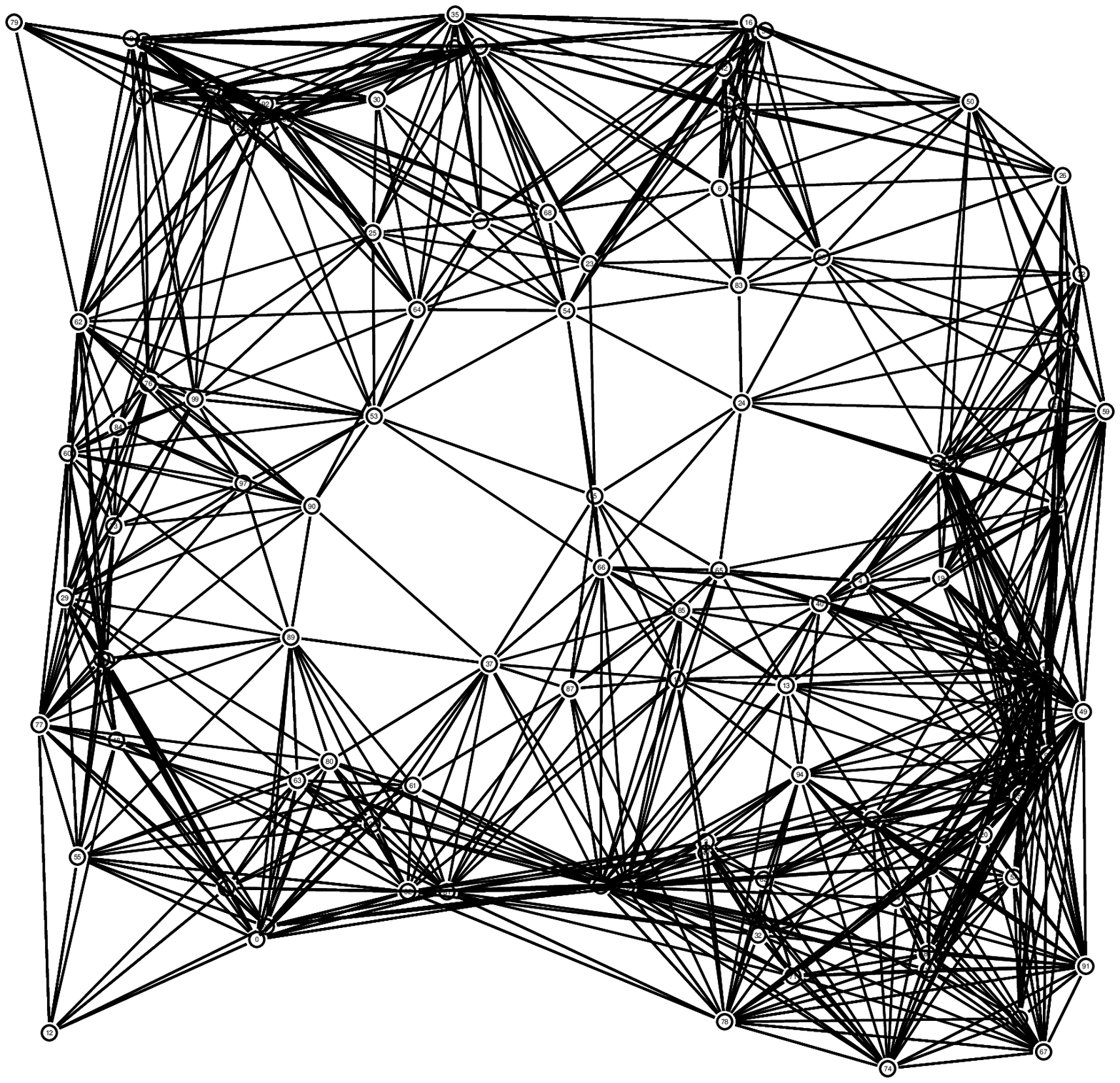}  \\
{\footnotesize (a) no topology control}  & 
{\footnotesize (b) $\alpha=2\pi/3$, basic algorithm}  \\
\epsfysize=5.0cm \epsffile{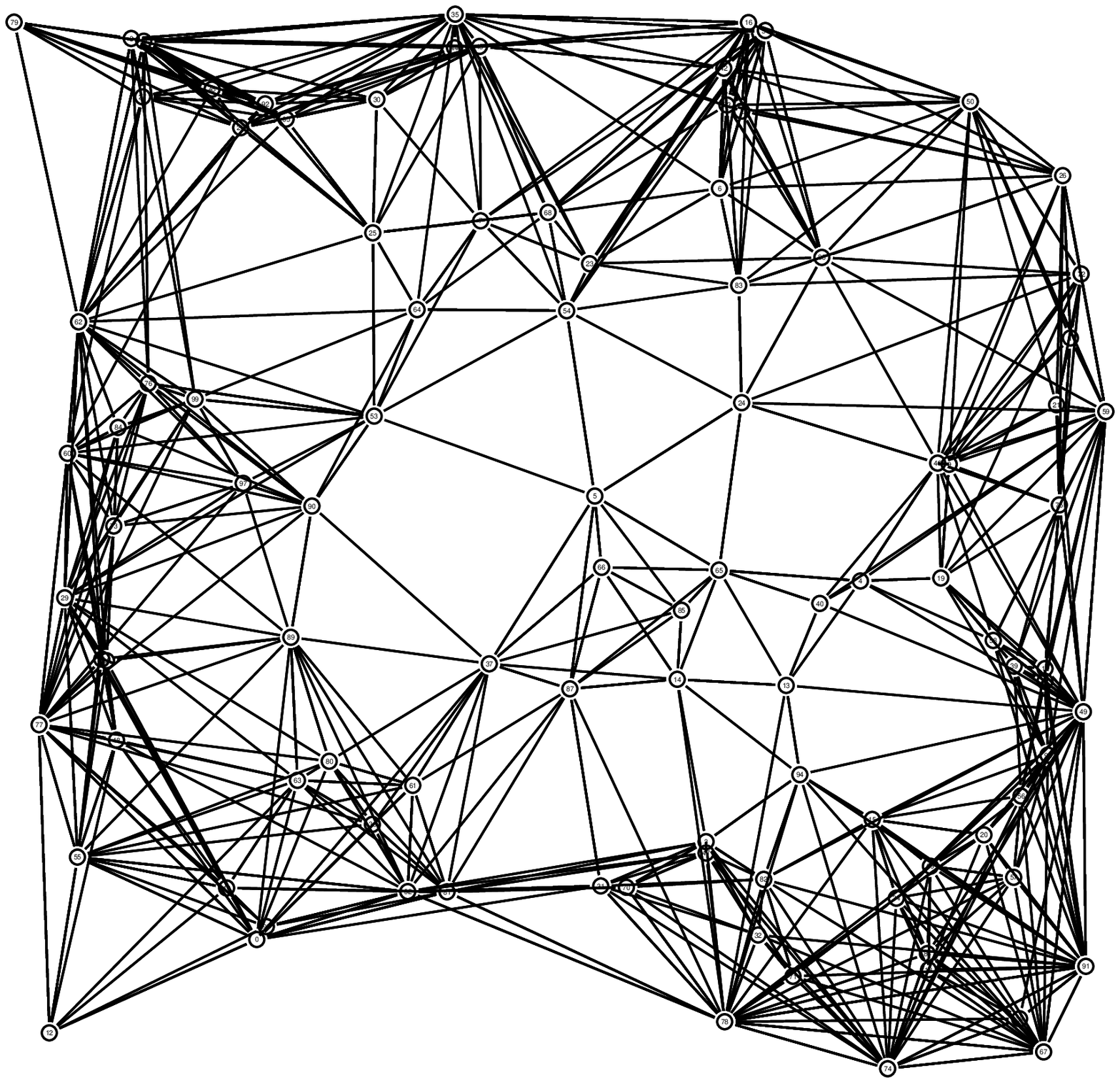} & 
\epsfysize=5.0cm \epsffile{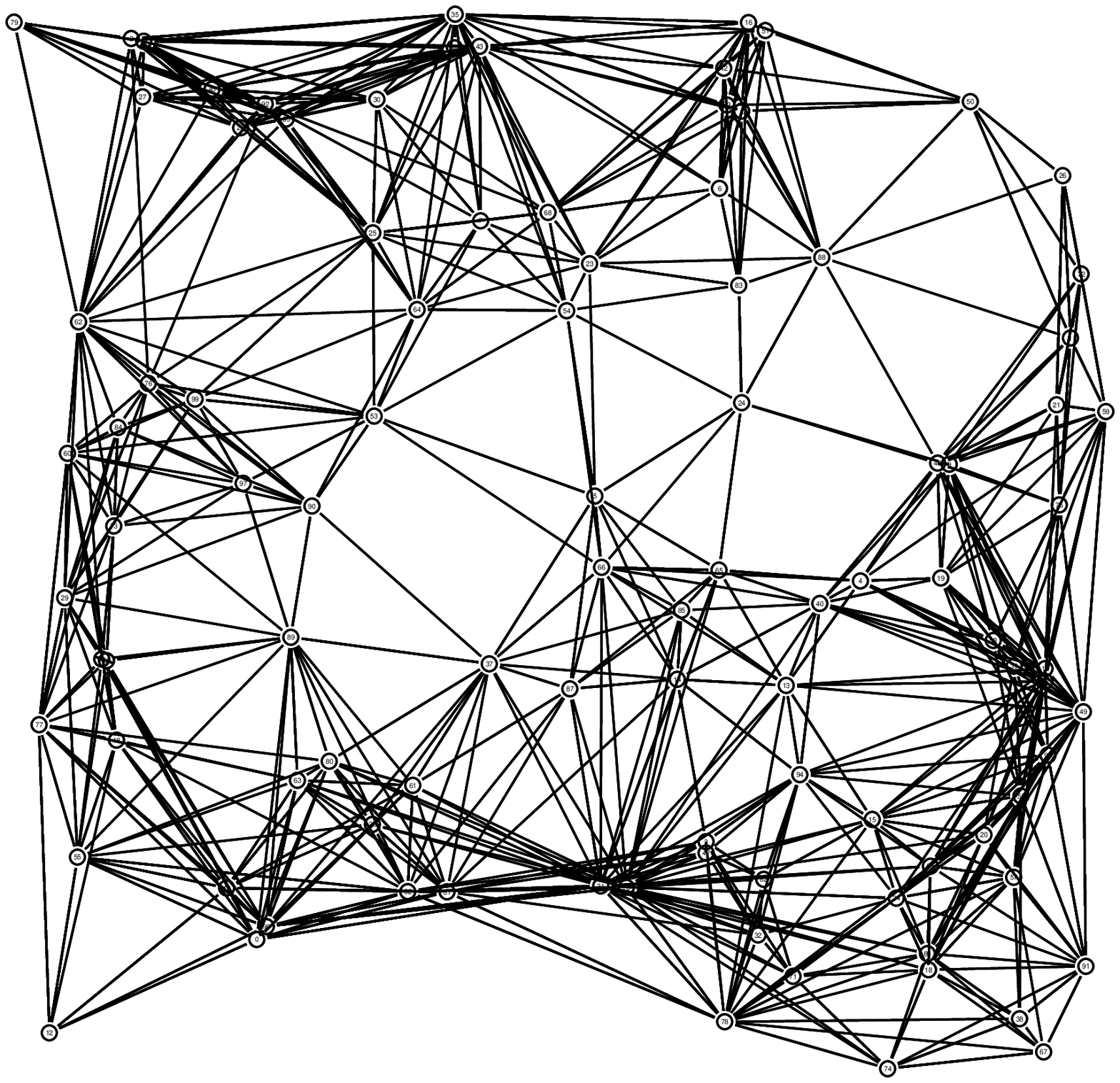} \\
{\footnotesize (c) $\alpha=5\pi/6$, basic algorithm} & 
{\footnotesize (d) $\alpha=2\pi/3$ with shrink-back} \\
\epsfysize=5.0cm \epsffile{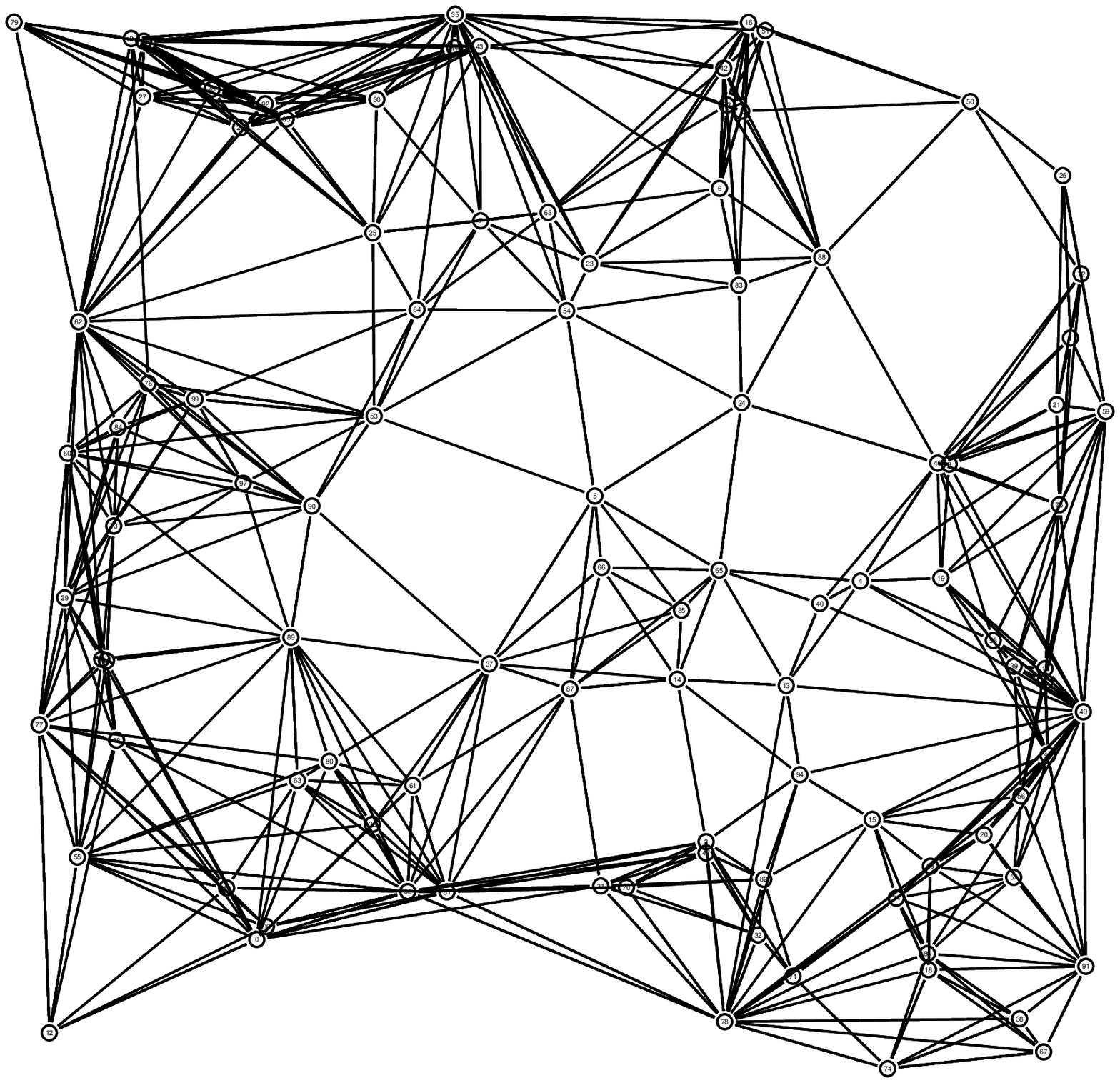} & 
\epsfysize=5.0cm \epsffile{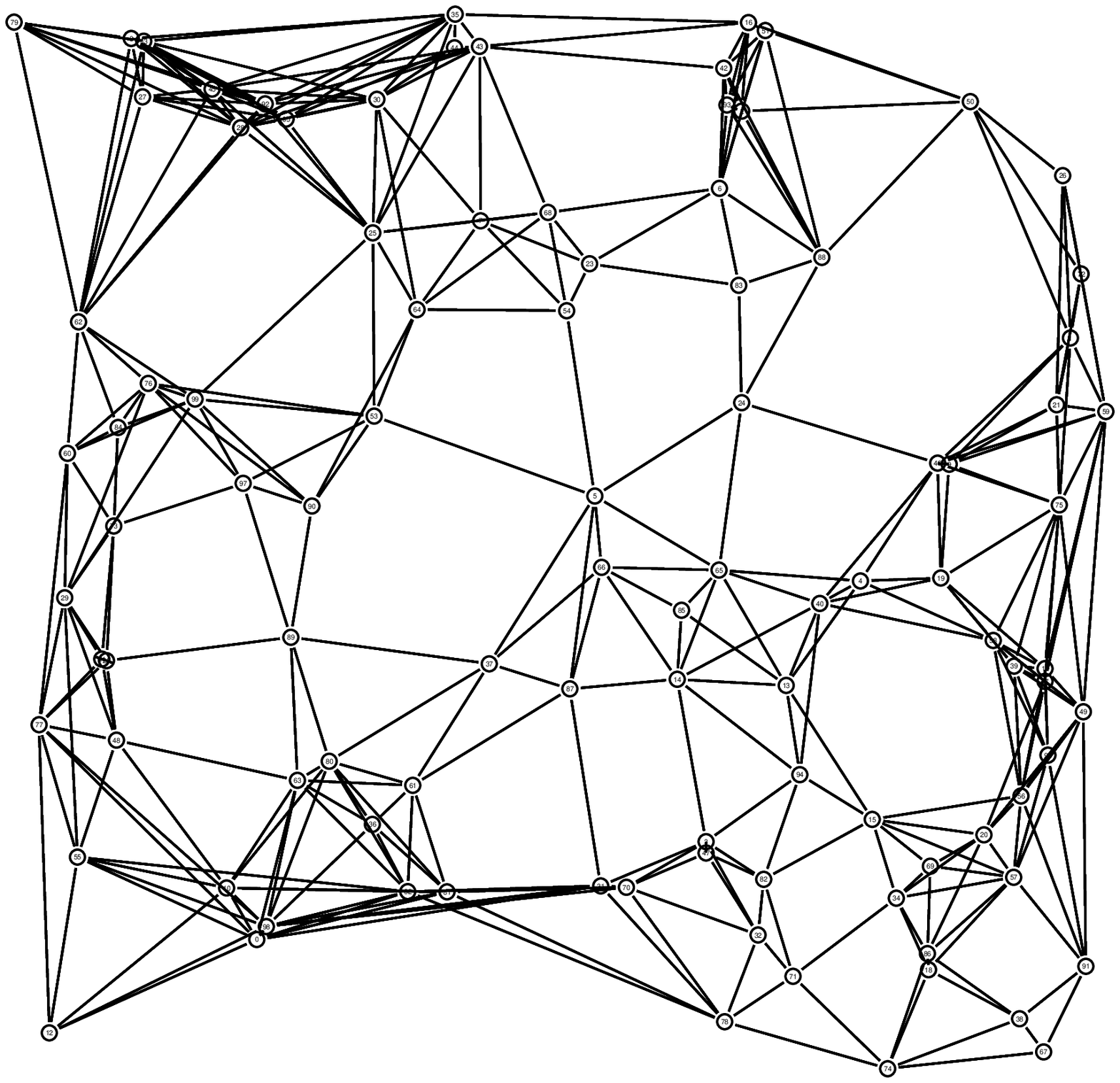 } \\
{\footnotesize (e) $\alpha=5\pi/6$ with shrink-back} & 
{\footnotesize (f) $\alpha=2\pi/3$ with shrink-back}  \\
& {\footnotesize and asymmetric edge removal} \\
\epsfysize=5.0cm \epsffile{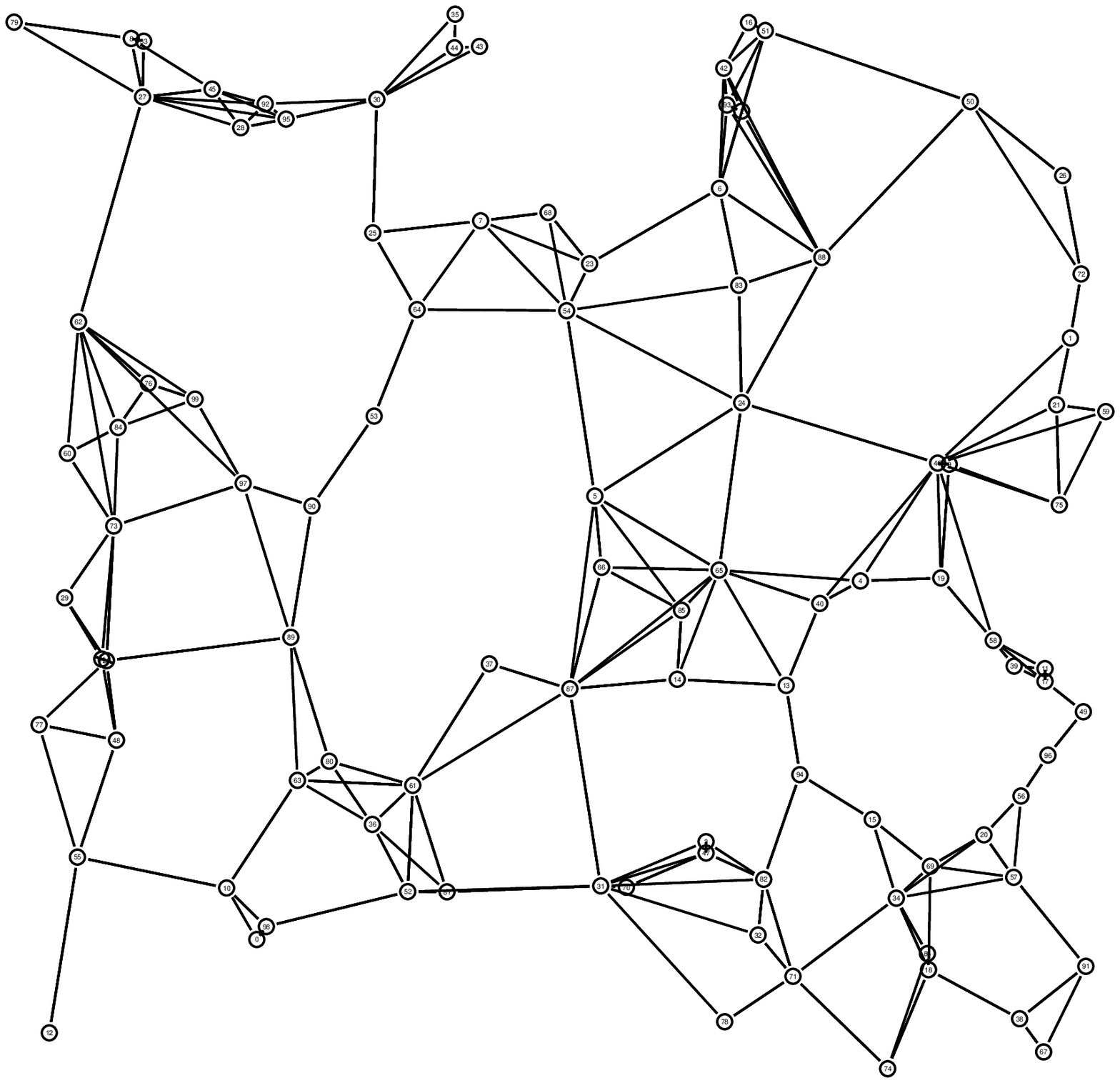} & 
\epsfysize=5.0cm \epsffile{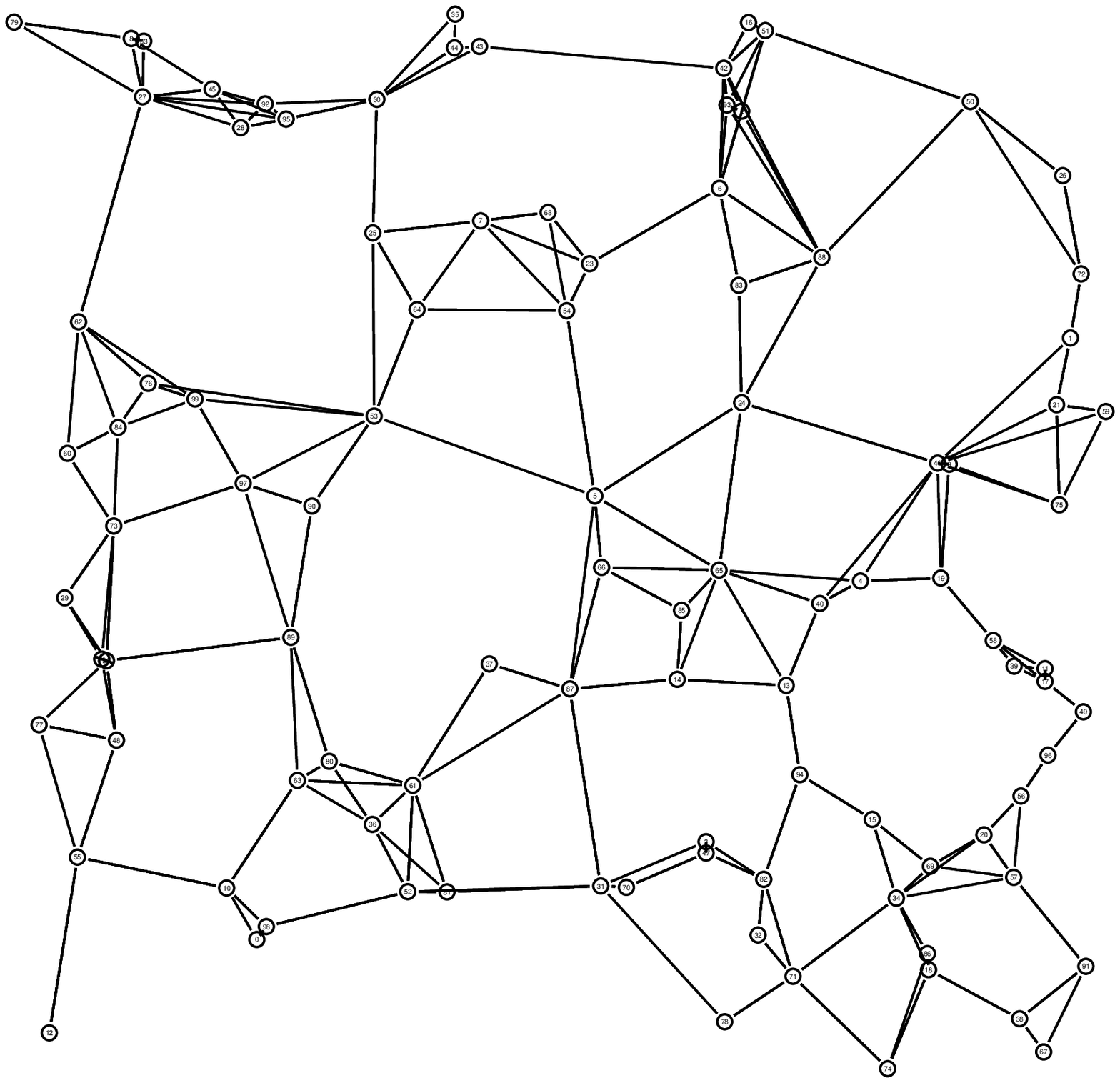} \\
{\footnotesize (g) $\alpha=5\pi/6$ with all applicable optimizations} &  
{\footnotesize (h) $\alpha=2\pi/3$ with all optimizations} \\
\end{tabular}
\end{center}
\caption{The network graphs as a result of different optimizations.
\label{fig-topo}
}
\end{figure*}

\section{Discussion}
\label{sec-conclusion}
\commentout{ 
We propose a simple distributed algorithm to guarantee a
connected topology with good performance. We prove the correctness of
the asynchronous algorithm.  
}

We have analyzed the distributed cone-based algorithm and proved that
$5\pi/6$ is a tight upper bound on the cone degree for the algorithm
to preserve connectivity.
We have also presented three optimizations 
to the basic algorithm%
---the shrink-back operation,
asymmetric edge removal, and pairwise edge removal---%
and proved that they improve performance while still preserving
connectivity. 
Finally, we showed that there is a tradeoff between using CBTC($\alpha$)
with 
$\alpha = 5\pi/6$ and $\alpha = 2\pi/3$, since using $\alpha = 2\pi/3$
allows an additional optimization, which can have a significant impact.
The algorithm extends easily to deal with reconfiguration and
asynchrony.  Most importantly, simulation results show that it is very
effective in reducing power demands.  

Reducing energy consumption has been viewed as perhaps the most
important design metric for topology control. 
There are two standard approaches to reducing energy consumption:
(1) reducing the 
transmission power of each node as much as possible;
(2) reducing the total energy consumption through the preservation of
minimum-energy paths in the underlying network.
These two approaches may conflict: reducing the transmission power
required by each node may not result in minimum-energy paths (see
\cite{ourInfocom01} for a discussion) or vice versa.  
Furthermore, there are other
metrics to consider, such as network throughput and network
lifetime. Reducing energy consumption tends to increase network
lifetime.  
(This is particularly true if the main reason that nodes die is due to
loss of battery power.)
However, there is no guarantee that it will. For example,
using minimum-energy paths for all communication
may result in hot spots and congestion,
which in turn may drain battery power and lead to network partition.
Using approach (1) in this case may do better (although there is no
guarantee).  
If topology control is not done
carefully, network throughput can be hurt. 
As we have already pointed out,
eliminating edges may result in more congestion and hence worse
throughput, even if it saves power in the short
run.  The right tradeoffs to make are very much application dependent.
We hope to explore these issues in more details in future work. 

{\small
\bibliographystyle{plain}
\bibliography{lilisbib}
}.

\end{document}